\documentclass[aps,prb,reprint,amsmath,amssymb,amsfonts,superscriptaddress]{revtex4-1}
\usepackage{graphicx,dcolumn,bm,color,multirow,microtype,amsmath,amssymb,rotating,subfigure}
\usepackage{physics}

\newcommand{\alert}[1]{\textcolor{black}{#1}}

\begin{document}

\title{Iterative stochastic subspace self-consistent field method}

\author{Pierre-Fran\c{c}ois Loos}
\thanks{Corresponding author}
\email{pf.loos@anu.edu.au}
\affiliation{Laboratoire de Chimie et Physique Quantiques, Universit\'e de Toulouse, CNRS, UPS, France}
\affiliation{Research School of Chemistry, Australian National University, Canberra, ACT 2601, Australia}
\author{Jean-Louis Rivail}
\email{jean-louis.rivail@univ-lorraine.fr}
\affiliation{Th\'eorie-Mod\'elisation-Simulation, Universit\'e de Lorraine, CNRS SRSMC UMR 7565, Vand{\oe}uvre-l\`es-Nancy F-54506, France}
\author{Xavier Assfeld}
\email{xavier.assfeld@univ-lorraine.fr}
\affiliation{Th\'eorie-Mod\'elisation-Simulation, Universit\'e de Lorraine, CNRS SRSMC UMR 7565, Vand{\oe}uvre-l\`es-Nancy F-54506, France}

\begin{abstract}
We propose a new self-consistent field (SCF) algorithm based on an iterative, partially stochastic \textit{``Divide \& Conquer''}-type approach. This new SCF algorithm is a simple variant of the usual SCF procedure and can be easily implemented in parallel. A detailed description of the algorithm is reported. We illustrate this new method on one-dimensional hydrogen chains and three-dimensional hydrogen clusters.
\end{abstract}


\maketitle

\section{Introduction}

Moore's law \cite{Moore65} predicts that the computer speed approximately doubles every two years. 
Therefore, we could expect that the system sizes accessible by conventional quantum chemistry \textit{ab initio} methods will keep growing rapidly in the next few decades. 
\alert{However, since several years, Moore's law is only fulfilled by an increasing number of cores per processor as CPU clock rate has stopped increasing.
Indeed, today hardware is designed in a multi-core manner and one needs to write softwares which support multi-threading to take full advantage of the hardware.}
Consequently, it becomes more and more important to develop parallel computer codes and algorithms that are capable for exploiting high-performance parallel computing environments. \cite{Nielsen08, deJong10, Yasuda08a, Yasuda08b, Ufimtsev08, Ufimtsev09a, Ufimtsev09b}

The computational bottleneck in Hartree-Fock (HF) \cite{SzaboBook} and density-functional theory (DFT) \cite{ParrBook} methods, in their conventional formulations, is the computation of the two-electron integrals which scales quartically with the number of basis functions $N$. 
Fortunately, this formal $O(N^4)$ scaling does reduce asymptotically to $O(N^2)$ if one takes into account only the numerically significant integrals \cite{Gill94a, Gill94b, Dyczmons73, Kussmann13, 3eRR} by using standard Cauchy-Schwarz screening \cite{Haser89} or distance-dependent screening techniques. \cite{Maurer12}

The two main steps in a self-consistent field (SCF) calculation are the formation of the Fock or Kohn-Sham (KS) matrix \cite{Challacombe97} and the subsequent determination of the molecular orbital (MO) coefficients. 
The work required for the formation of the Fock or KS matrix can be done very efficiently by exploiting the local nature of chemistry. 
For example, the formation of the Coulomb part of the Fock or KS matrix scales linearly with the system size using the continuous fast-multipole method developed by White and coworkers. \cite{White94a, White94b} 
Linear scaling can also be reached for the computation of the exact exchange matrix \cite{Schwegler96} and exchange-correlation potentials (see Refs.~\onlinecite{Goedecker99, Kussmann13}, and references therein).

In the conventional formulation of HF and DFT methods, the MO coefficients are obtained by diagonalization of the Fock or KS matrix, respectively. This step scales cubically with the system size, \cite{SzaboBook, ParrBook} and may therefore become the time dominating step for large molecules.

The present method aims at reducing the $O(N^3)$ cost of the matrix diagonalization by partitioning the MOs of the system into subsets and performing smaller diagonalizations in these subsets. 
The main advantage of the method is that each smaller diagonalization can be perform on a distinct core. 
The method is a MO-based variant of the SCF algorithm based on an iterative, partially stochastic \textit{``Divide \& Conquer''} strategy. \cite{Yang91, Yang95} 
This new algorithm that we have named \textit{iterative stochastic subspace} SCF (I3SCF) is a simple modification of the usual SCF procedure and can be easily implemented in parallel. 
Although major modifications and refinements have been made, the present method is inspired by the \textit{ab initio} LSCF method \cite{Assfeld96, Fornili06, Loos07, Monari13} used in QM/MM methods. \cite{Moreau04, Ferre02, Loos08, Loos09} 

We would like to mention here that the present investigation is a \textit{``proof of principle''} study.
Obviously, a more detailed study would be necessary to fully understand the advantages and limitations of the present algorithm, as well as calculations on larger systems.
We will report on this in the future.

The present article is organized as follows: in Sec.~\ref{sec:theory}, we give a detailed derivation of each step of the I3SCF algorithm, while illustrative examples are given in Sec.~\ref{sec:applications}.
Atomic units are used throughout.

\begin{figure}
	\includegraphics[width=0.8\linewidth]{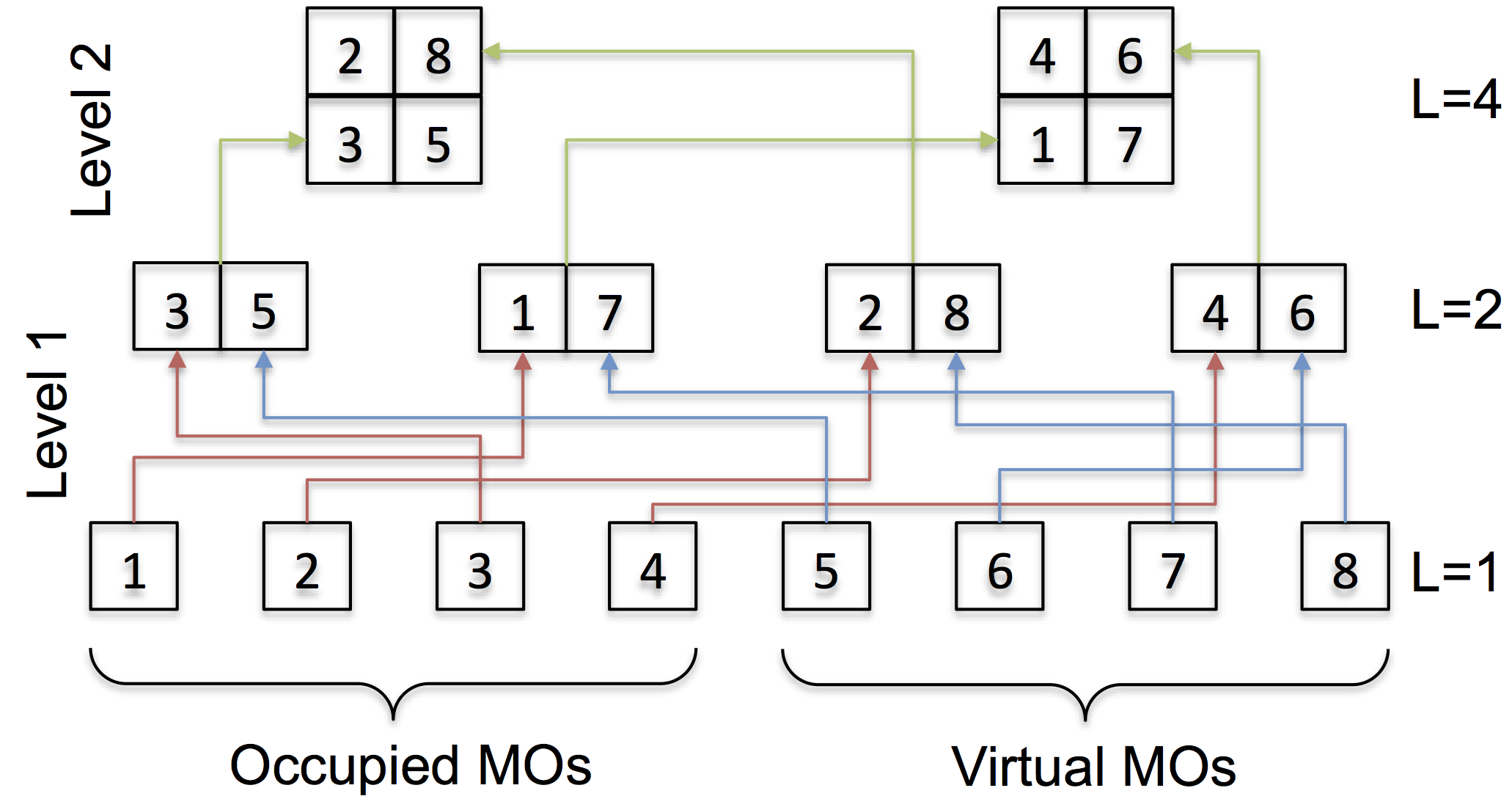}
\caption{
\label{fig:partition}
Schematic representation of the stochastic partition with $N=8$ and $K=2$.}
\end{figure}

\begin{figure*}
	\includegraphics[width=\linewidth]{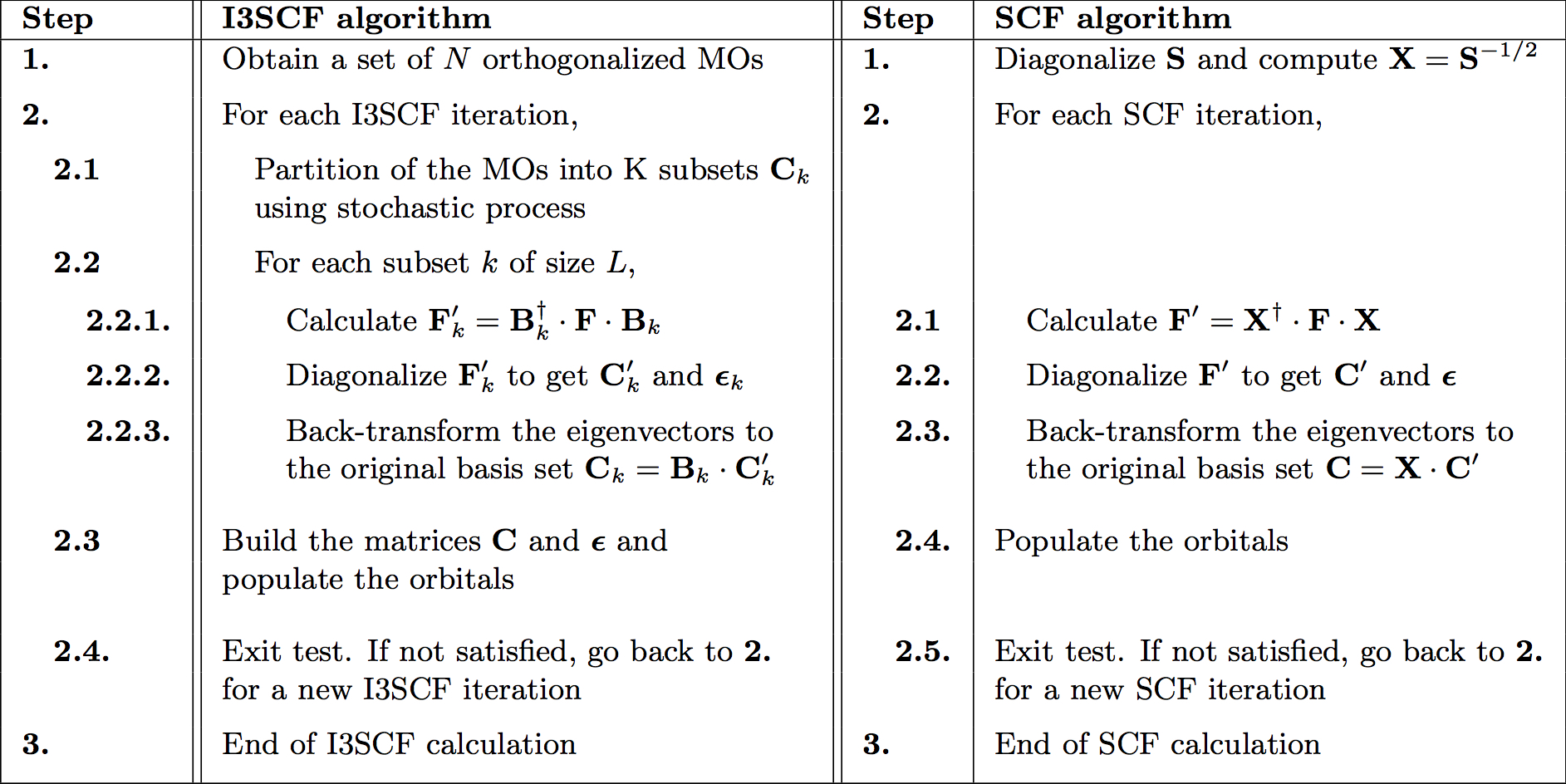}
\caption{
\label{fig:algo}
Main steps in the I3SCF and SCF algorithms.}
\end{figure*}

\section{\label{sec:theory}Theory}

\subsection{Stochastic partition}

The purpose of the stochastic process is to partition the MOs $\qty{ \ket{\psi_p} }_{1 \le p \le N}$ into $K$ subsets of size $L$ at each SCF cycle. 
For sake of simplicity, we assume that the $K$ subsets have the same dimension $L$. 
However, the present process can be easily generalized for subsets of different sizes (see below). 
I3SCF calculations using $K$ subsets are labelled I3SCF[$K$]. 
The atomic orbital (AO) coefficients of the $k$th subset are gathered in the matrix $\mathbf{C}_k$ of dimension $N \times L$. For $K=1$ (i.e.~$L=N$), the present algorithm is equivalent to the conventional SCF procedure. 
First, to illustrate the stochastic partition procedure, we set $N=2^m$, where $m \in \mathbb{N}^*$. 
The general case is discussed below.

The first and crudest level of partition (level 1 in Fig.~\ref{fig:partition}) consists in creating occupied-virtual MO pairs by \textit{randomly} choosing an occupied MO $\psi_i$ and associating it with the virtual MO $\psi_a$ which has the largest coefficient
\begin{equation}
	\mathbf{f}_{ia} = \norm*{ \mathbf{c}_i^\dag \cdot \mathbf{F} \cdot \mathbf{c}_a }_\mathcal{F},
\end{equation}
where $\mathbf{F}$ is the Fock matrix, the vectors $\mathbf{c}_i$ and  $\mathbf{c}_a$ contain the AO coefficients of $\psi_i$ and $\psi_a$, and 
\begin{equation}
	\norm*{ \mathbf{Z} }_\mathcal{F} = \sqrt{\Tr(\mathbf{Z}^{\dag} \cdot \mathbf{Z})}
\end{equation}
is the Frobenius norm.

This maximizes the probability of decreasing the energy by mixing the occupied-virtual MO pairs with large off-diagonal Fock elements. 
We anticipate that these ``level 1'' calculations will exhibit the slowest convergence rate due to the restricted number of MOs per subset ($L=2$). 
At level 1, the stochastic nature of the algorithm is due to the random choice of the occupied MOs. 
It is interesting to note that the level 1 partition has some similarities with the well-known Jacobi sweep technique. \cite{NumericalRecipes} 

The next step of the stochastic process (level 2 in Fig.~\ref{fig:partition}) associates two occupied-virtual MO pairs to create larger subsets (if required). 
This is done by \textit{randomly} picking a subset $k$ and associating it with the subset $k^{\prime}$ having the largest Frobenius norm \cite{GolubBook}
\begin{equation}
\label{fock}
	\mathbf{f}_{k k^{\prime}} = \norm*{\mathbf{C}_k^\dag \cdot \mathbf{F} \cdot \mathbf{C}_{k^\prime}}_\mathcal{F}.
\end{equation}
At level 2, the stochastic nature of the algorithm is due to the random choice of the subset $k$.

This process is repeated until the required number of subsets is created. 
This stochastic partition is schematically illustrated in Fig.~\ref{fig:partition}. 
In the case where $N \neq 2^m$, the level 1 partition can be modified in order to build the target number of subsets. 
For example, two virtual MOs can be associated with one occupied MO to create triplets. 
Then, the level 2 partition associates these triplets to create larger subsets.

\begin{figure}
	\includegraphics[width=0.8\linewidth]{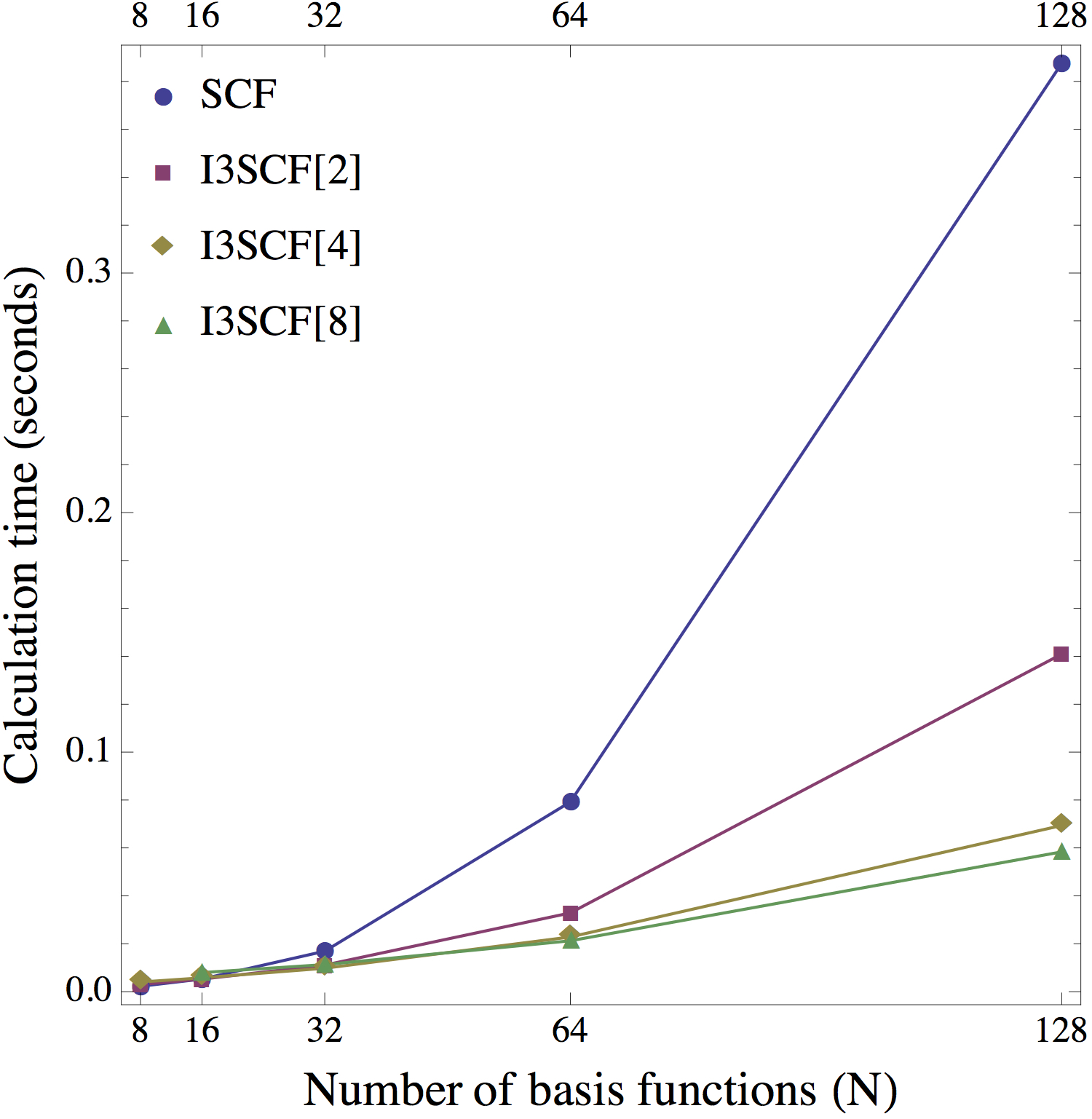}
\caption{
\label{fig:CPU}
Calculation time (in second) for performing step \textbf{2.2.2.}~of the I3SCF[$K$] algorithm as a function of $N$ for various $K$ values. 
The calculation time of step \textbf{2.2.}~in the conventional SCF algorithm is also reported for comparison.}
\end{figure}

\begin{figure*}
	\includegraphics[width=0.32\textwidth]{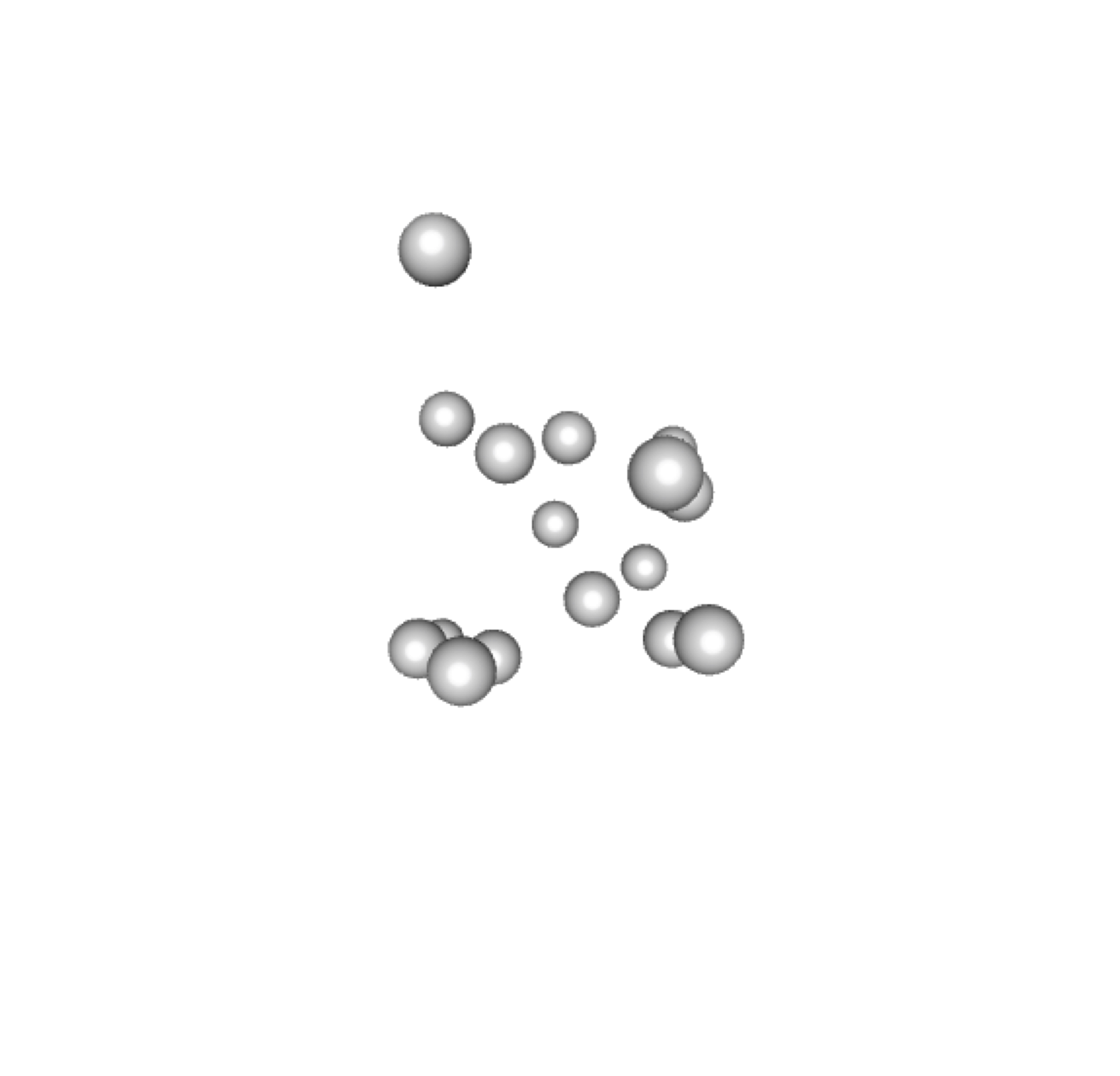}
	\includegraphics[width=0.32\textwidth]{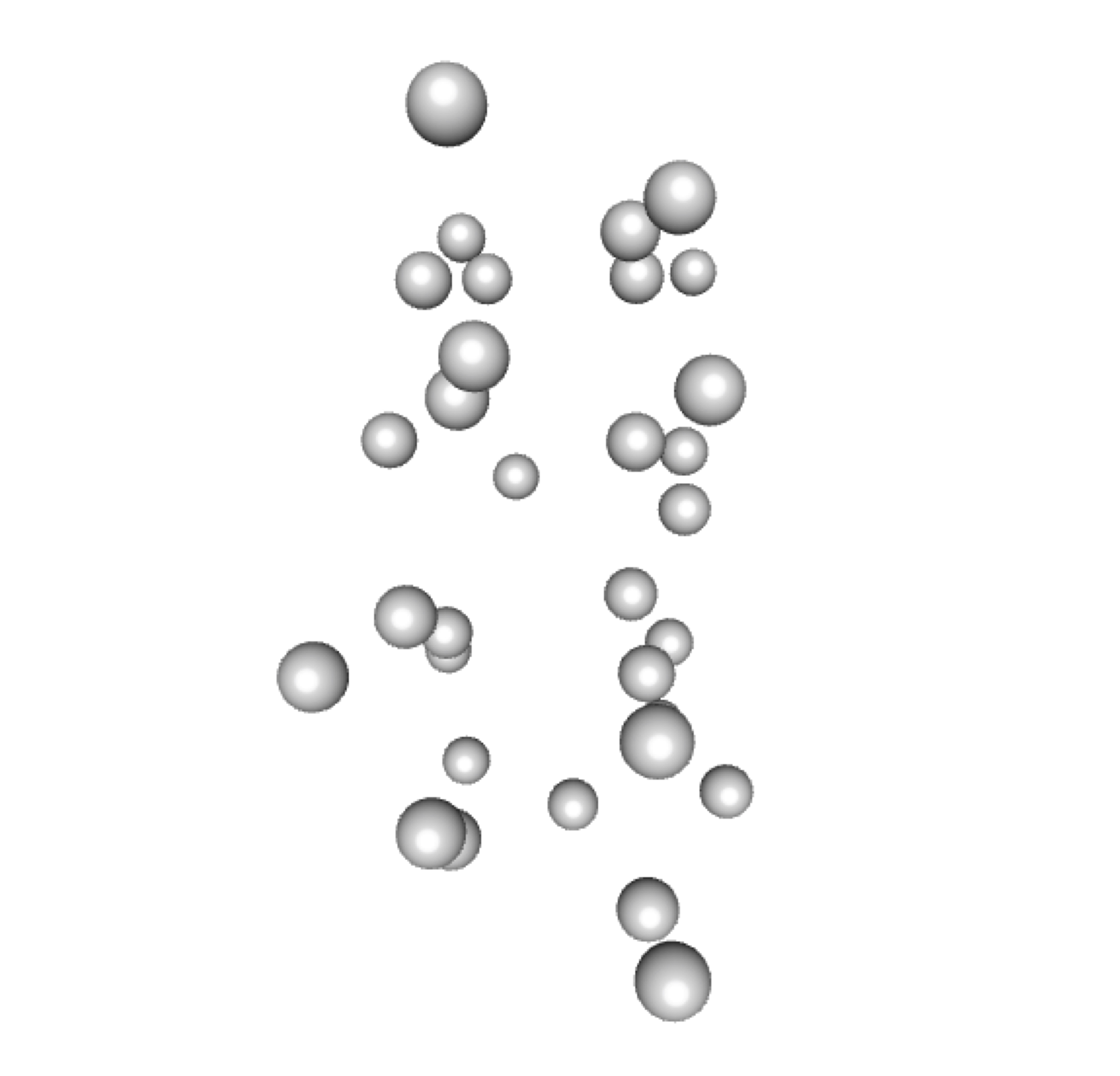}
	\includegraphics[width=0.32\textwidth]{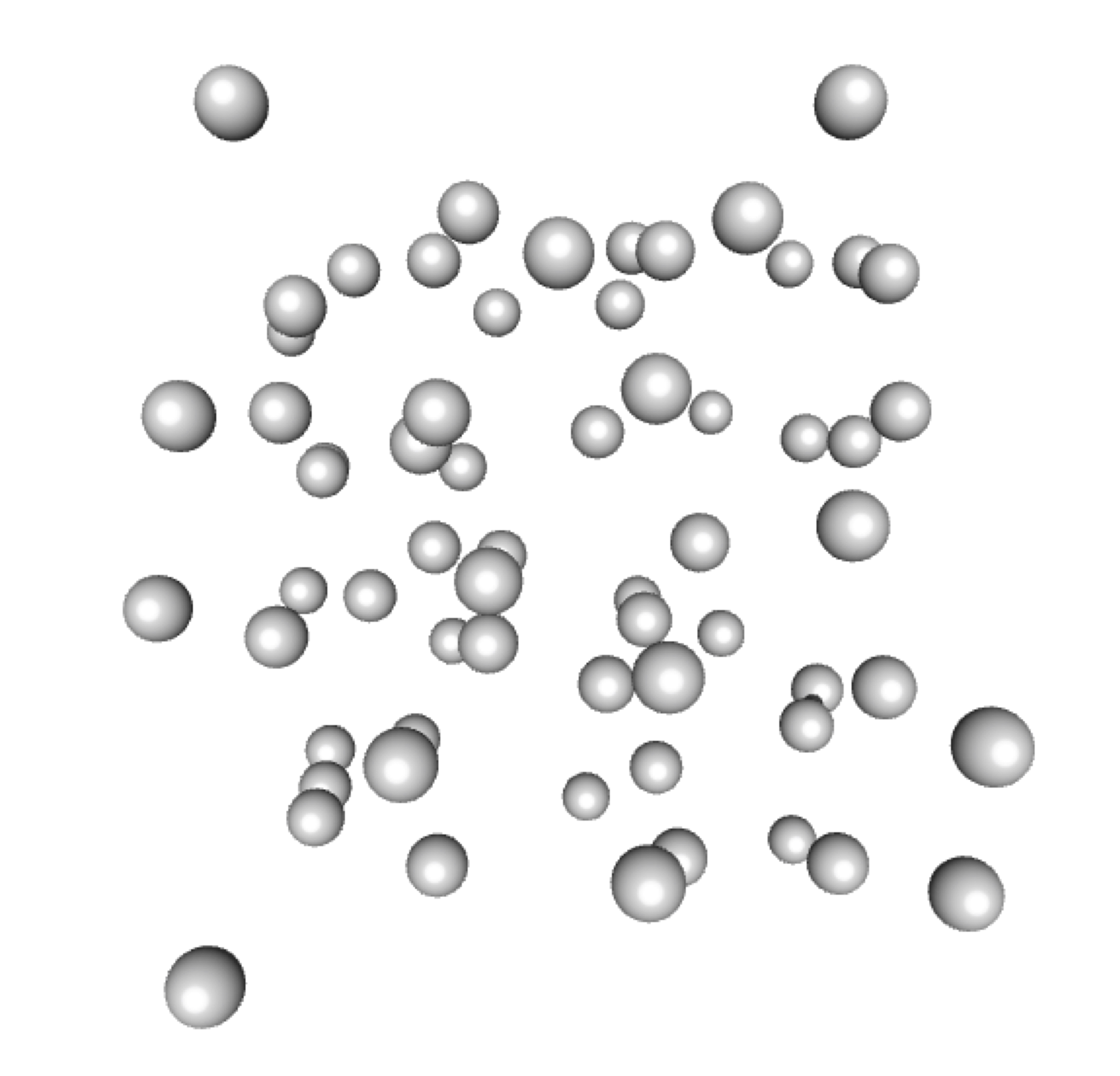}
\caption{
\label{fig:cluster}
\alert{Structure of the three-dimensional hydrogen clusters: H$_{16}$ (left), H$_{32}$ (center) and H$_{64}$ (right).}}
\end{figure*}

\begin{figure}
	\includegraphics[width=0.8\linewidth]{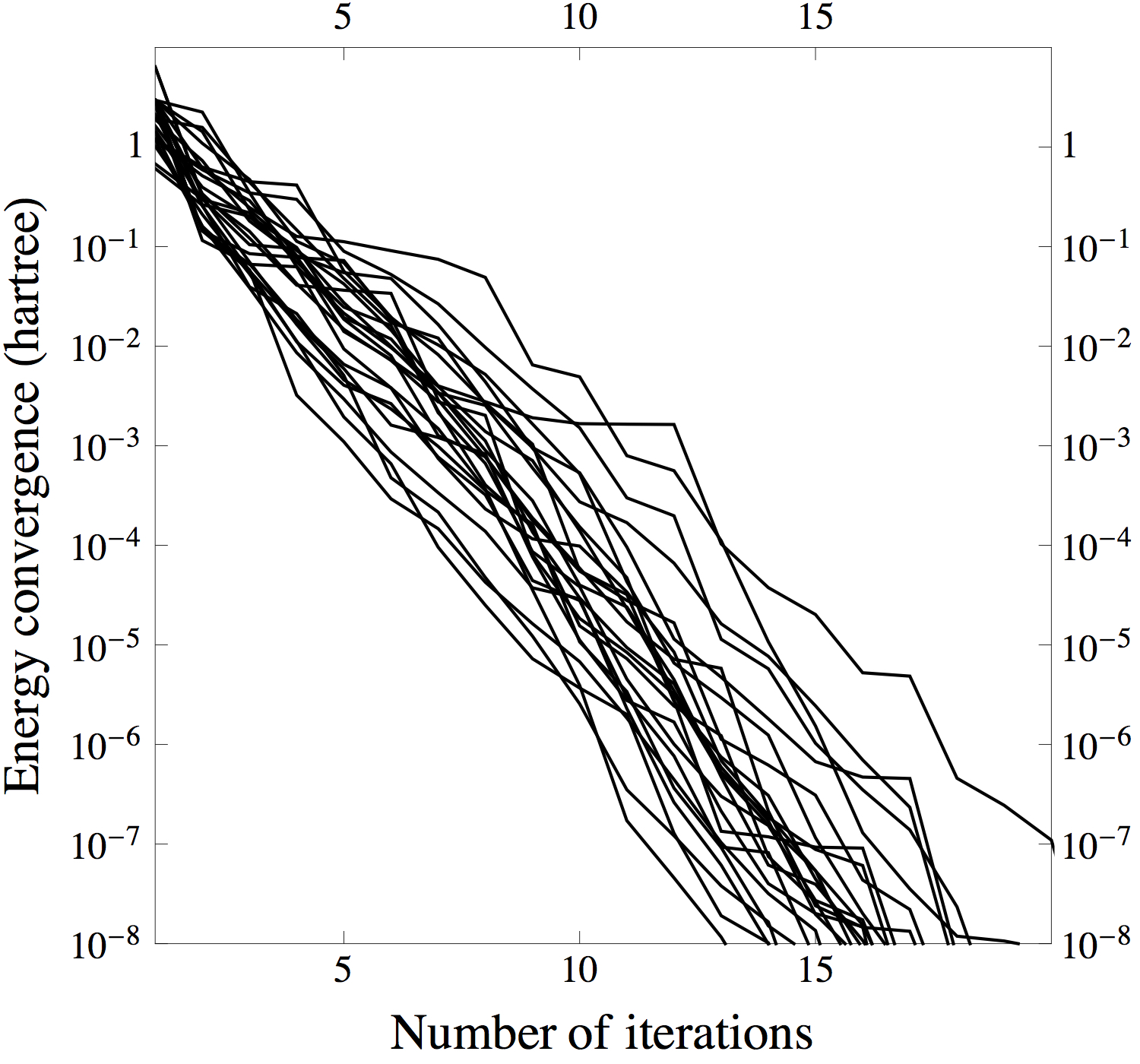}
\caption{
\label{fig:seed}
Energy convergence (in hartree) of the H$_{16}$ chain for 25 different random seeds using the I3SCF algorithm with $L=8$.}
\end{figure}

\subsection{Diagonalization}

The next step consists in computing and diagonalizing the transformed Fock matrix 
\begin{equation}
	\mathbf{F}^{\prime}_k = \mathbf{B}_k^{\dag} \cdot \mathbf{F} \cdot \mathbf{B}_k
\end{equation}
of size $L \times L$ to obtain the eigenvectors $\mathbf{C}^{\prime}_k$ and eigenvalues $\boldsymbol{\epsilon}_k$. 
We use the orthogonalization matrix \cite{SzaboBook}
\begin{equation}
	\label{B-C}
	\mathbf{B}_k = \mathbf{C}_k,
\end{equation}
\alert{where $\mathbf{C}_k$ in Eq.~\eqref{B-C} refers to the MO coefficients from the previous iteration.}
The matrix $\mathbf{B}_k$ fulfills the desired orthogonality condition, i.e.~
\begin{equation}
	\mathbf{B}_k^{\dag} \cdot \mathbf{S} \cdot \mathbf{B}_k = \mathbf{I}
\end{equation}
(where $\mathbf{I}$ is the identity matrix), which must be satisfied if the transformed orbitals are to form an orthogonal set.
Note that the procedure requires an orthogonal set of $N$ MOs as initial guess. 

The eigenvectors are then back-transformed to the original basis \alert{to obtain a new set of MO coefficients:}
\begin{equation}
	\mathbf{C}_k = \mathbf{B}_k \cdot \mathbf{C}_k^{\prime}.
\end{equation}

After this diagonalization step has been performed on each subset, we form the matrices 
\begin{align}
	\mathbf{C} & = \bigcup_{k=1}^K \mathbf{C}_k,
	&
	\boldsymbol{\epsilon} & =  \bigcup_{k=1}^K \boldsymbol{\epsilon}_k,
\end{align}
where the union of two matrices $\mathbf{Z}_1$ and $\mathbf{Z}_2$ returns a matrix with the rows of $\mathbf{Z}_1$ followed by the rows of $\mathbf{Z}_2$. 
The orbitals are then populated using the Aufbau principle. \cite{SzaboBook} 

The I3SCF algorithm is summarized in Fig.~\ref{fig:algo}.
For comparison, we also report the main steps of the usual SCF procedure. 
Loop \textbf{2.2.}~over the $K$ subsets can be efficiently parallelized as each subset can be allocated to a single core. 
The main advantage of the I3SCF algorithm is that the expensive $O(N^3)$ diagonalization step of the ``total'' Fock matrix is reduced to smaller diagonalizations. Because each of these diagonalizations only cost $O(L^3)$ (step \textbf{2.2.2.}), the total computational cost (i.e.~over the $K$ subsets) is reduced to $K \times O(L^3) = O(N L^2)$ for the I3SCF algorithm. 
This is a significant reduction in cost compared to the $O(N^3)$ diagonalization cost of the usual SCF algorithm.

We illustrate this point in Fig.~\ref{fig:CPU}, where we compare the calculation time as a function of $N$ for step \textbf{2.2}~in the conventional SCF algorithm and for step \textbf{2.2.2}~in the I3SCF algorithm. 	
As one can see, for any value of $K$, the \textit{``Divide \& Conquer''} diagonalization strategy of the I3SCF algorithm eventually becomes faster than the ``full'' diagonalization required by the conventional SCF, with a cost growing near-linearly with $N$. 
In comparison, the usual SCF method exhibits a steep increase of the calculation time.

\begin{figure*}
	\includegraphics[height=0.39\textheight]{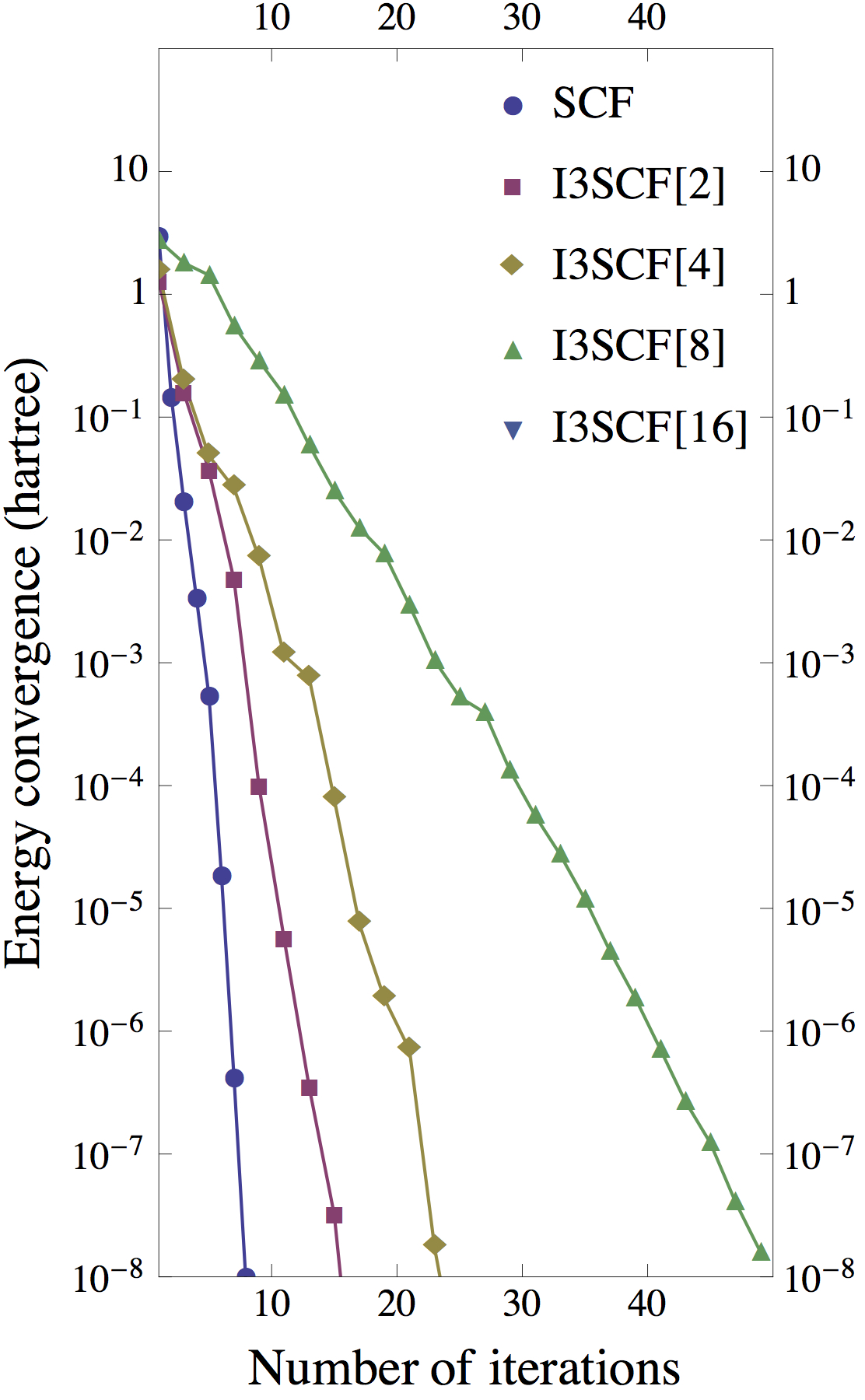}
	\includegraphics[height=0.39\textheight]{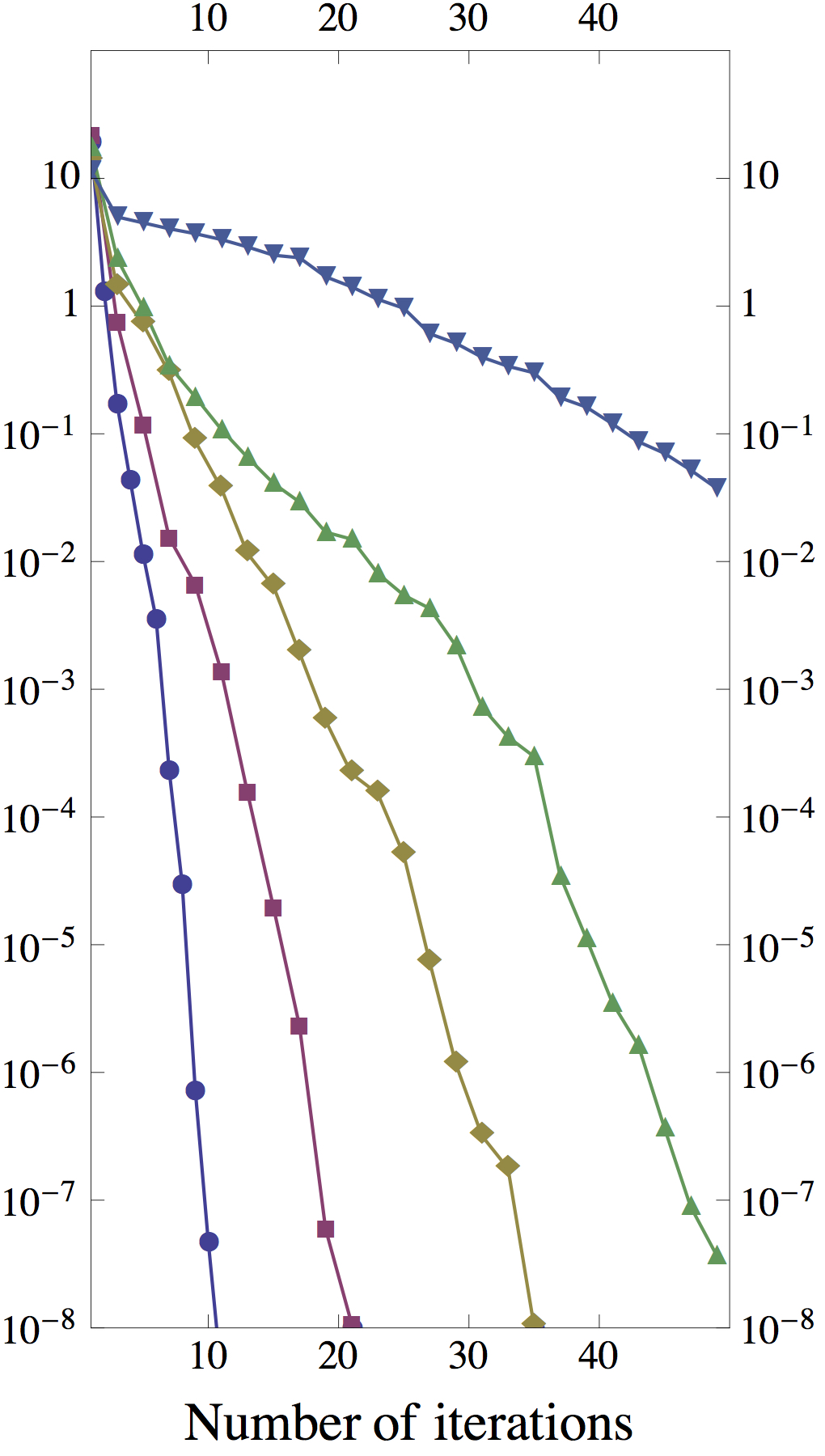}
	\includegraphics[height=0.39\textheight]{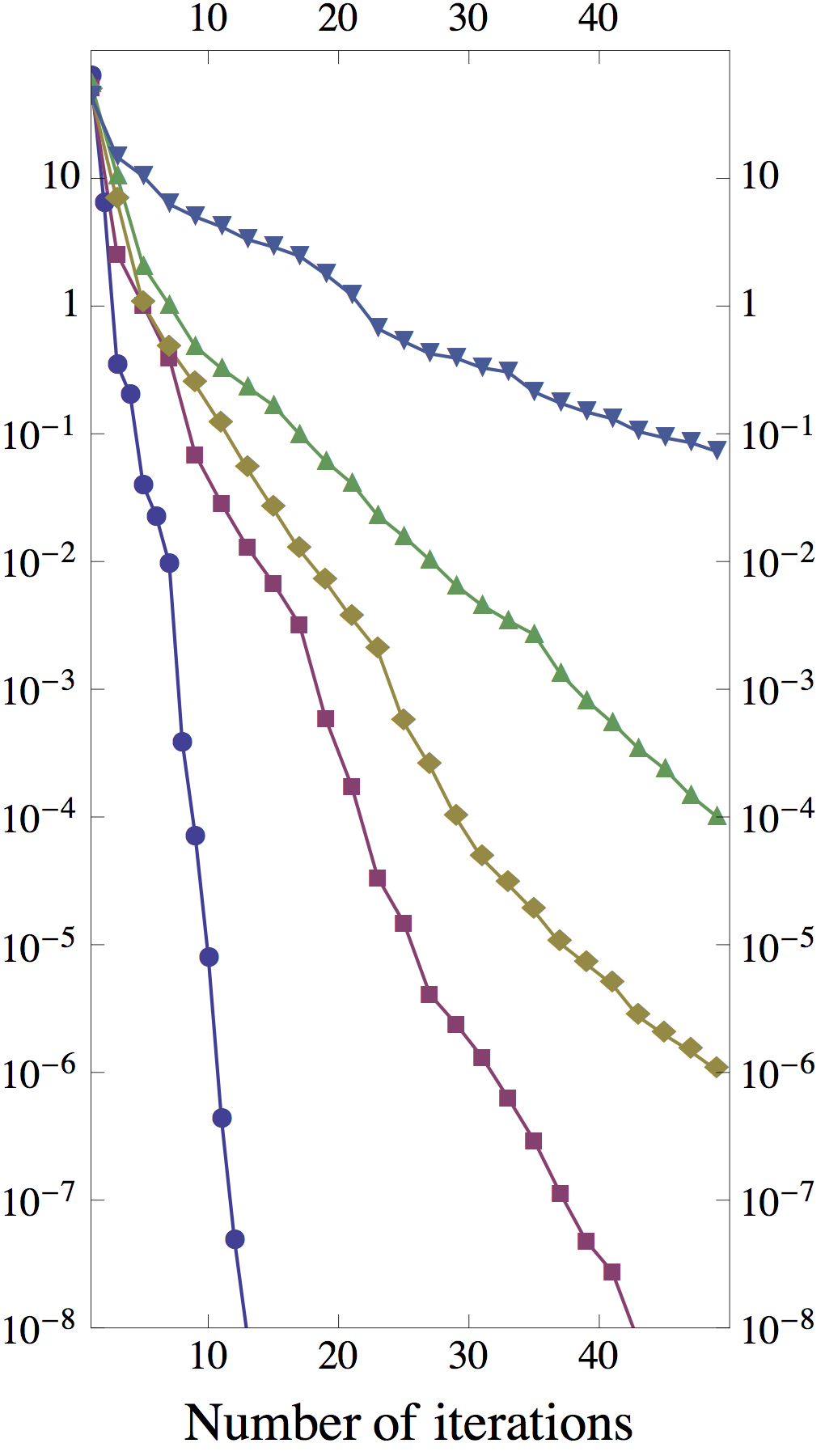}
\caption{
\label{fig:1D-H}
Energy convergence (in hartree) of one-dimensional hydrogen chains: H$_{16}$ (left), H$_{32}$ (center) and H$_{64}$ (right).} 
\end{figure*}

\section{\label{sec:applications}Applications}

\subsection{Computational details}

To illustrate the advantages and disadvantages of the I3SCF algorithm, we have computed the restricted HF ground-state energies of linear chains of $n$ equally-spaced hydrogen atoms for $n=16$, $32$ and $64$ with the same number of spin-up and spin-down electrons (closed-shell systems).
The distance $R_\text{HH}$ between neighboring atoms has been set to $1.8$ bohr. \cite{Tsuchimochi09} 
These systems have recently attracted considerable interest due to their strong correlation character and metal-insulator transition. \cite{Tsuchimochi09, Sinitskiy10, SanchoGarcia11, Stella11} 

To test the performance of the I3SCF algorithm for non-symmetric structures, we have also computed the restricted HF ground-state energies of three-dimensional hydrogen clusters for $n=16$, $32$ and $64$. 
These clusters are obtained by considering  the regular $2 \negmedspace \times \negmedspace 2 \negmedspace \times \negmedspace 4$, $2 \negmedspace \times \negmedspace 4 \negmedspace \times \negmedspace 4$ and $4 \negmedspace \times \negmedspace 4 \negmedspace \times \negmedspace 4$ lattices with $R_\text{HH}=1.8$ bohr and applying a random Gaussian displacement of standard deviation $\sigma = R_\text{HH}/4$ in each direction. 
\alert{These are represented in Fig.~\ref{fig:cluster}, and} the cartesian coordinates of the corresponding structures are reported as supplementary material.

The conventional SCF algorithm as well as the I3SCF algorithm have been implemented in \textsc{Mathematica}. \cite{Mathematica} 
The AOs basis consists of a single Gaussian function of exponent $0.4$, which has been obtained in order to reproduce the STO-6G basis. \cite{Hehre69, Hehre70} 
All the calculations uses the core Hamiltonian as a guess Fock matrix, and Pulay's DIIS method \cite{Pulay80, Pulay82} is applied to accelerate convergence. 

\begin{figure*}
	\includegraphics[height=0.39\textheight]{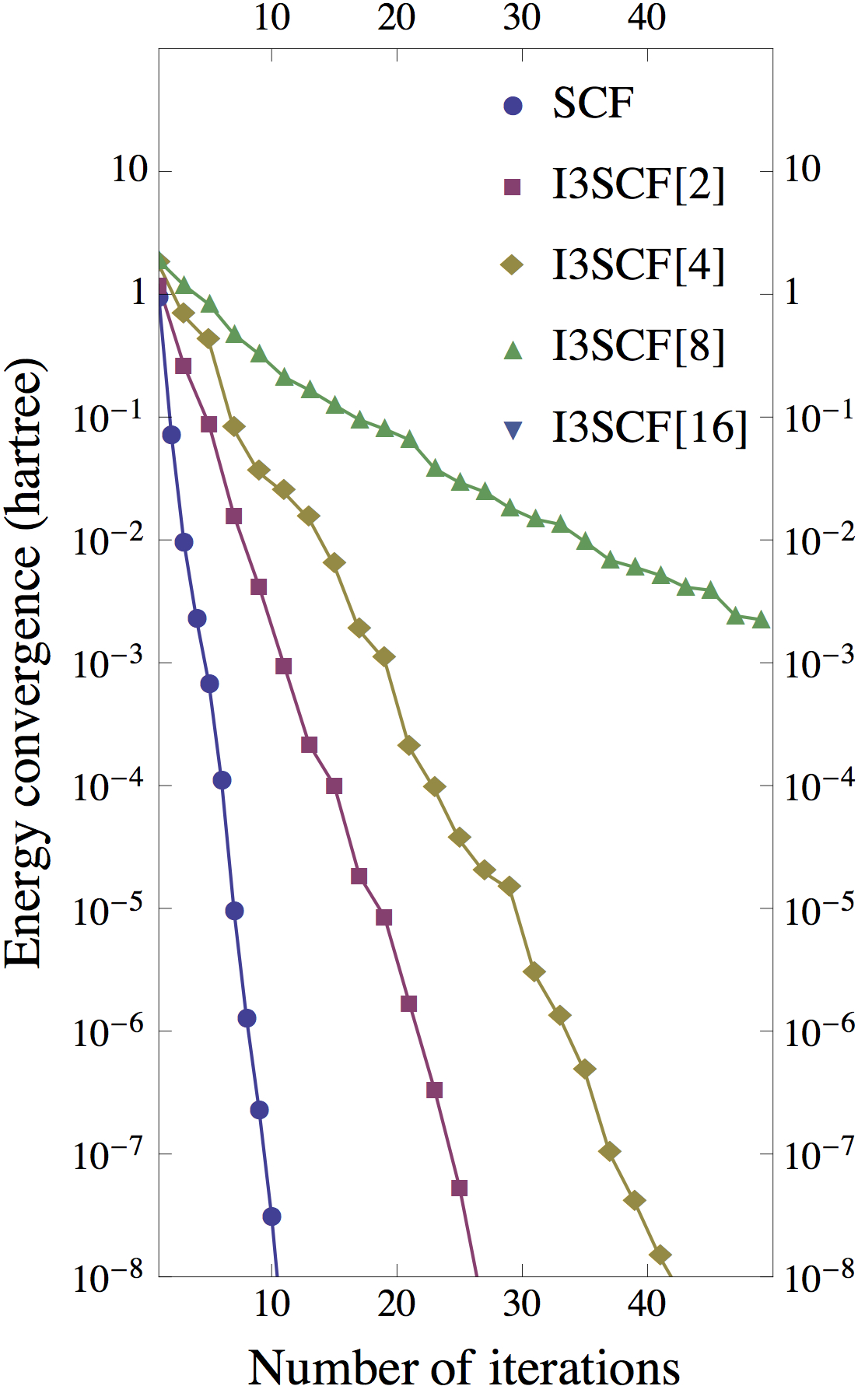}
	\includegraphics[height=0.39\textheight]{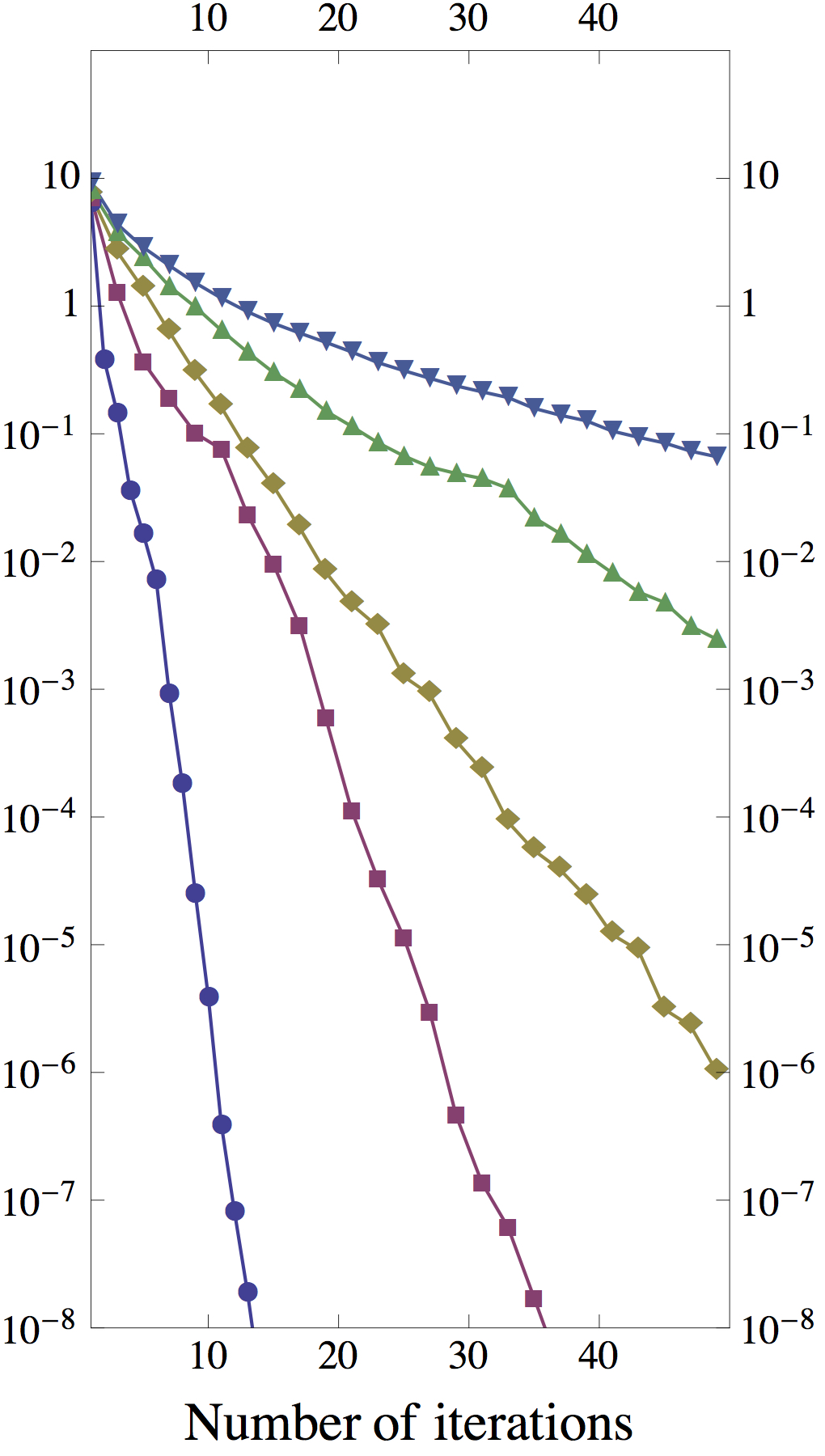}
	\includegraphics[height=0.39\textheight]{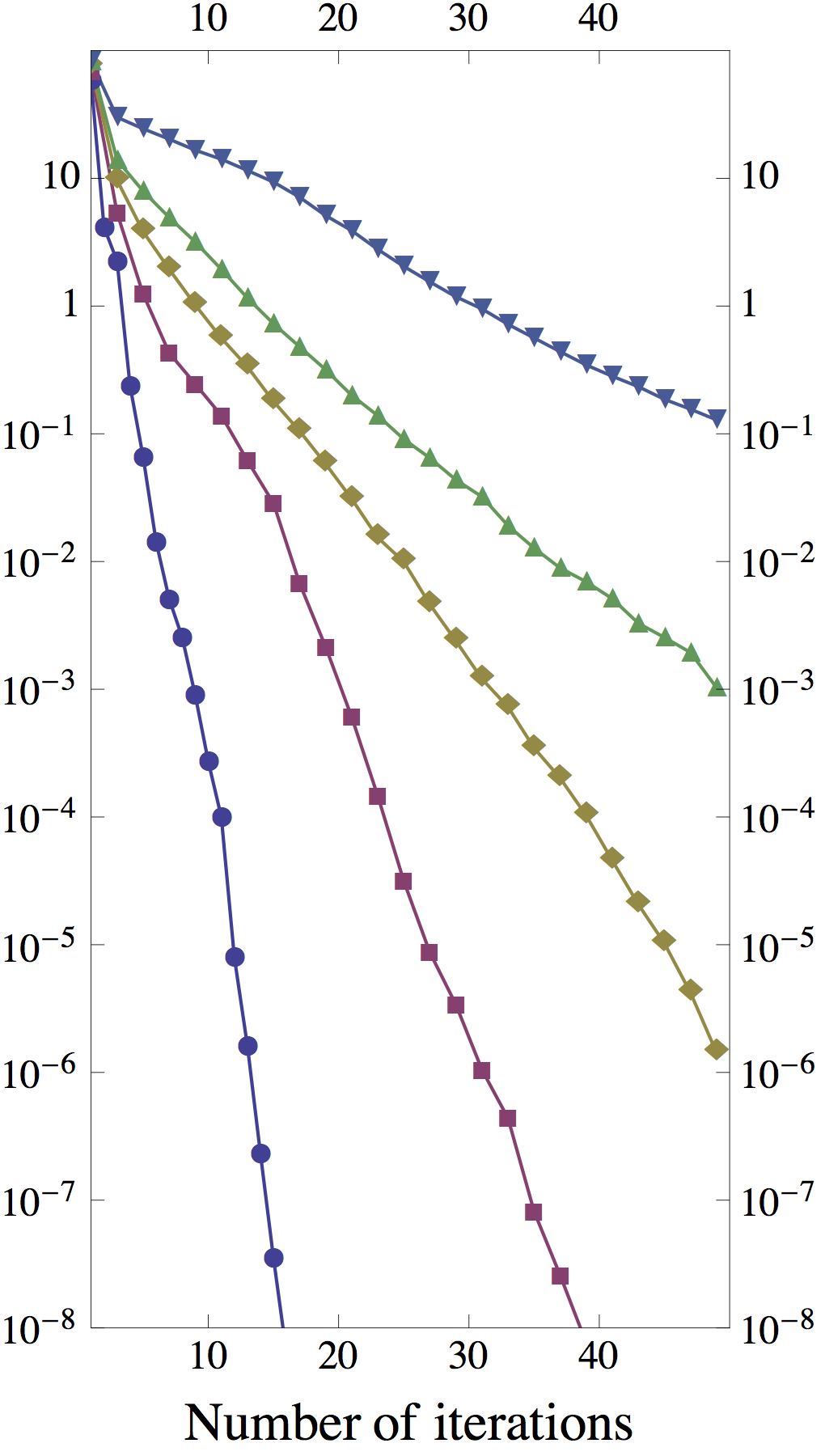}
\caption{
\label{fig:3D-H}
Energy convergence (in hartree) of three-dimensional hydrogen clusters: H$_{16}$ (left), H$_{32}$ (center) and H$_{64}$ (right). }
\end{figure*}

\subsection{Hydrogen chains and clusters}

First, we have tested the dependence of the I3SCF algorithm with respect to the stochastic partitioning by performing I3SCF[2] calculations on the H$_{16}$ chain using 25 different random seeds. 
In Fig.~\ref{fig:seed}, we have plotted the energy convergence (in hartree) defined as the difference between the energy at a given iteration and the converged energy. 
As one can see, the final result is largely independent of the choice of the random seed.

The energy convergence of the H$_{16}$,  H$_{32}$ and H$_{64}$ chains is represented in Fig.~\ref{fig:1D-H}. 
To reach a given energy accuracy, the I3SCF algorithm requires more iterations than the conventional SCF method. 
For a given value of $K$, the number of iterations seems to increase linearly with the system size.\footnote{In all the cases studied in the work, we have observed a smooth convergence of the SCF energy. Therefore, the curves reported in Figs.~\ref{fig:1D-H} and \ref{fig:3D-H} can be easily extrapolated to the desired convergence threshold.}
The same behavior is observed for the SCF algorithm. 
This means that the present method does not shift the cost of each diagonalization step to the number of diagonalization steps.
Because each diagonalization can be done on a distinct core, a parallel implementation of the I3SCF algorithm could be competitive with the conventional SCF algorithm, especially for large systems. 
The computation cost of the I3SCF method could be further reduced by employing localized orbitals \cite{Guo11} and sparse algebra routines. \cite{Challacombe99, TewarsonBook}

For small $L$, the I3SCF algorithm needs a larger number of iterations to achieve the desired convergence threshold, due to the small number of MOs per subset. 
For example, the I3SCF[8] calculation on H$_{16}$ has only two MOs per subset. 
However, for any value of $K$, the I3SCF algorithm converges reliably to the SCF limit.

The same conclusions can be drawn for the three-dimensional hydrogen clusters, as shown in Fig.~\ref{fig:3D-H}. 
Even for non-symmetric systems in higher dimensions, the I3SCF algorithm converges reliably to the SCF limit. 
Moreover, we observe that, for a given value of $K$, the number of iterations required to reach a given energy accuracy is roughly independent of the system size. 
Again, the same observation can be made for the SCF algorithm.
Also, we note that, the convergence is getting slower when the number of subsets $K$ increases.
However, this could be fixed by using localized orbitals that will impose a spatial constraint in the stochastic partition in order to avoid having weakly interacting MOs in the same subset. 
We will investigate this possibility in a forthcoming paper.
In some cases the overall computation gain can be rather limited (as typically in I3SCF[2] calculations).
It is therefore important to find the best compromise between the size of the Fock matrices to diagonalize and the level of stochastic partition (see Fig.~\ref{fig:partition}).

\section{Conclusion}

In this \alert{article}, we have described and studied a new SCF algorithm that we have dubbed \textit{iterative stochastic subspace} SCF (I3SCF). 
This new method, which is a simple variant of the usual SCF algorithm, is based on a \textit{``Divide \& Conquer''} strategy which partitions the MOs of the system into subsets. 
It can be parallelized efficiently on modern parallel computers. 
The I3SCF algorithm has been tested on one-dimensional and three-dimensional hydrogen systems, for which it has shown promising performances. 
We hope to report results for larger molecules of biological interest in the near future.

\begin{acknowledgments}
PFL thanks the NCI National Facility for a generous grant of supercomputer time, and the Australian Research Council for funding (Grant DP140104071) and a Discovery Early Career Researcher Award (Grant DE130101441). 
XA would like to thank the Research School of Chemistry of the Australian National University for a visiting fellowship during the construction of this manuscript.
\end{acknowledgments}

\end{document}